\documentclass[twocolumn, tighten, times, twocolappendix]{aastex631}

\begin{document}

\title{Mixed origins: strong natal kicks for some black holes and none for others}

\author[0000-0002-1386-0603]{Pranav Nagarajan}
\affiliation{Department of Astronomy, California Institute of Technology, 1200 E. California Blvd., Pasadena, CA 91125, USA}

\author[0000-0002-6871-1752]{Kareem El-Badry}
\affiliation{Department of Astronomy, California Institute of Technology, 1200 E. California Blvd., Pasadena, CA 91125, USA}



\begin{abstract}

Using stellar kinematic data from {\it Gaia} DR3, we revisit constraints on black hole (BH) natal kicks from observed accreting and detached BH binaries. We compare the space velocities and Galactic orbits of a sample of 12 BHs in the Galactic disk with well-constrained distances to their local stellar populations, for which we obtain proper motions and radial velocities from {\it Gaia} DR3. Compared to most previous studies, we infer lower minimum kick velocities, because our modeling accounts for the fact that most BH binaries are old and have likely been kinematically heated by processes other than kicks. Nevertheless, we find that half of the BHs have at least weak evidence for a kick, being kinematically hotter than at least 68\% of their local stellar populations. At least 4 BHs are kinematically hotter than 90\% of their local stellar populations, suggesting they were born with kicks of $\gtrsim 100$ km s$^{-1}$. On the other hand, 6 BHs have kinematics typical of their local populations, disfavoring kicks of $\gtrsim 50$ km s$^{-1}$. For two BHs, V404 Cyg and VFTS 243, there is strong independent evidence for a very weak kick $\lesssim 10$ km s$^{-1}$. Our analysis implies that while some BHs must form with very weak kicks, it would be wrong to conclude that {\it most} BHs do, particularly given that selection biases favor weak kicks. Although the uncertainties on most individual BHs' kicks are still too large to assess whether the kick distribution is bimodal, the data are consistent with a scenario where some BHs form by direct collapse and receive weak kicks, and others form in supernovae and receive strong kicks.

\end{abstract}



\section{Introduction} \label{sec:intro}

When massive stars die, they can deliver an impulse to their compact object remnants known as a natal kick. These kicks arise due to recoil from asymmetric mass loss and anisotropic neutrino emission during the supernova explosion, with the latter mechanism predicted to be subdominant for neutron stars (NSs) but important for black holes (BHs) (see \citealt{lai_neutron_2004} for a review, and \citealt{burrows_theory_2024} or \citealt{janka_interplay_2024} for recent 3D simulations of core-collapse supernovae). In binaries, even a symmetric supernova explosion can lead to a recoil in the orbital plane known as a ``Blauuw'' kick \citep{blaauw_1961}. For large kicks or significant amounts of mass loss, binaries can become unbound during a supernova \citep{hills_effects_1983, brandt_effects_1995}.

It has long been appreciated that NSs have high space velocities relative to their progenitors \citep{lyne_high_1994}. A statistical study of pulsar proper motions by \citet{hobbs_statistical_2005} found that their 3D space velocities (derived via deconvolution) are well fit by a Maxwellian with $\sigma = 265$ km s$^{-1}$, though other kinematic studies have argued for a bimodal velocity distribution \citep[e.g.,][]{igoshev_observed_2020}. Indeed, the high retention fraction of NSs in globular clusters \citep{pfahl_comprehensive_2002} and observed populations of wide NS binaries \citep{phinney_verbunt_1991, pfahl_new_2002, podsiadlowski_effects_2004, knigge_two_2011, khargaria_2012, fortin_constraints_2022, odoherty_observationally_2023, van_der_wateren_psr_2024, el-badry_19_2024, nagarajan_symbiotic_2024} provide strong evidence that some NSs are born with small natal kicks.

While NS kicks have been studied extensively, BH kicks are less well understood. Studying the kinematics of the population of BHs in the Milky Way is key to constraining the magnitude of the natal kicks that these BHs receive. Unfortunately, while isolated NSs can be observed as radio pulsars, isolated BHs can only be studied through rare microlensing events. Only one secure BH microlensing event is known to date \citep{lam_isolated_2022, sahu_isolated_2022, mroz_2022, lam_2023}, and its observed properties do not tightly constrain its natal kick \citep{andrews_constraining_2022}. Instead, BH natal kicks can be inferred using binaries that feature a luminous companion (LC) orbiting a BH. In these systems, the kinematics of the LC or the radio emission from a compact jet can be used to infer the 3D space velocity of the BH. Fortunately, more than 20 BHs in X-ray binaries have been dynamically confirmed to date \citep{remillard_x-ray_2006, corral-santana_blackcat_2016}. In addition, astrometry from the \textit{Gaia} mission \citep{gaia_collaboration_gaia_2016} has recently enabled the discovery of three non-interacting BHs in wide orbits \citep{el-badry_sun-like_2023, el-badry_red_2023, gaia_collaboration_discovery_2024}. Dormant BH candidates are also suspected on dynamical grounds in X-ray quiet binaries in the Milky Way disk \citep{mahy_identifying_2022}, the globular cluster NGC 3201 \citep{giesers_detached_2018, giesers_2019}, and the Large Magellanic Cloud \citep{shenar_x-ray-quiet_2022}. 

Because strong natal kicks disrupt binaries, we expect kicks inferred from BH binaries to be biased towards weak natal kicks relative to the entire BH population. Similarly, kicks inferred from isolated BHs are expected to be biased toward strong natal kicks. However, since most microlensing searches focus on long-duration events, they will be biased towards low BH velocities and thus small natal kicks.

Individual studies of BH natal kicks have been performed on several BH binaries, including Cyg X-1 \citep{nelemans_constraints_1999, wong_understanding_2012}, GRO J1655-40 \citep{brandt_high_1995, nelemans_constraints_1999, mirabel_2002, willems_understanding_2005}, XTE J1118+480 \citep{mirabel_high-velocity_2001, gualandris_has_2005, fragos_understanding_2009}, GRS 1915+105 \citep{dhawan_kinematics_2007}, MAXI J1305-704 \citep{kimball_black_2023}, H 1705-250 \citep{dashwoodbrown_2024}, and VFTS 243 \citep{Stevance2023, vigna-gomez_constraints_2024}. Overall, results from these works have been mixed, with some studies finding evidence for significant kicks and others finding none. 

There have also been several studies analyzing the kinematics of the known BH population as a whole \citep[e.g.,][]{repetto_investigating_2012, repetto_constraining_2015, mandel_estimates_2016, repetto_2017, atri_potential_2019, zhao_evidence_2023}.  Many of these studies have used the peculiar velocity and/or scale height of BH X-ray binaries as a proxy for the natal kick, implicitly assuming that BHs born with no kicks should remain on kinematically cold, thin disk orbits. This approach may overestimate natal kicks for BHs in old stellar populations, which could potentially have large peculiar velocities and be found far from the disk midplane even if they formed with no kick \citep[e.g.,][]{wielen_1977, nordstrom_2004, mackereth_2019}. 

Several recent developments have provided new insight into natal kicks likely to have been imparted on individual BHs, motivating us to revisit inference of BH kicks from the known BH population. First, two BH-LC systems --- V404 Cyg and VFTS 243 --- have recently been shown unequivocally to have been born with very weak kicks. In the case of V404 Cyg, the system is actually part of a hierarchical triple, and only a natal kick $\lesssim 5$ km s$^{-1}$ allows the system to remain bound \citep{burdge_2024}. Meanwhile, for VFTS 243, which has not yet been tidally synchronized, the near-circular orbit implies a natal kick $\lesssim 10$ km s$^{-1}$ \citep{Stevance2023, vigna-gomez_constraints_2024}. Second, on the other end of the spectrum, the recently discovered low-mass X-ray binary Swift J1727.8-162 has one of the highest observed peculiar velocities for a dynamically confirmed BH ($207 \pm 7$ km s$^{-1}$), implying that the BH was born with a strong kick \citep{matasanchez_2024}. Third, precise parallaxes and proper motions from \textit{Gaia}'s 3rd data release \citep[``DR3";][]{gaia_collaboration_gaia_2023} enable a kinematic analysis of BH-LC systems for which such measurements were not previously available in the literature. Fourth, radial velocities from \textit{Gaia} DR3 provide 3D motions for large comparison samples of stars in the vicinity of known BHs, making it possible to probe the kinematics of BHs' local stellar populations.

In this study, we investigate BH natal kicks using a sample of 12 BH-LC systems with distance, proper motion, and radial velocity measurements from the literature. In Section~\ref{sec:sample}, we discuss our sample selection process and visualize the distribution of the selected binaries in the Milky Way. In Section~\ref{sec:analysis}, we analyze the natal kicks of these systems, accounting for the velocity dispersion of their local stellar populations. We calculate Toomre diagrams and integrated Galactic orbits for each system. In Section~\ref{sec:discussion}, we discuss the significance of our results, and compare them against the literature. We also discuss implications for models of binary evolution and core-collapse supernovae. Finally, in Section~\ref{sec:conclusion}, we summarize our results and consider avenues for future work. Appendix~\ref{appendix:co_comparison} compares the natal kicks of BHs studied in this work to the natal kicks of NSs with low-mass companions.

\section{Sample Selection} \label{sec:sample}

We compile literature distances, proper motions, and center-of-mass radial velocities for systems hosting luminous stars orbiting dynamically confirmed BHs in the Milky Way and Large Magellanic Cloud (LMC). In general, we adopt the most recent and precise distance and center-of-mass RV measurements available in the literature. The techniques used to estimate distances include optical and radio parallax measurements, spectral modeling (e.g.\ assuming the absolute magnitude of a main sequence star of the best-fit spectral type), ellipsoidal variability modeling (i.e.\ assuming that the donor is a main sequence star that fills its Roche lobe), and observations of the proper motions of jet ejecta \citep[for a summary of distance estimation techniques in the literature, see][]{jonker_2004}. In cases where the distance comes from a {\it Gaia} parallax, we adopt the distances from \citet{bailerjones_2021}, which include an exponentially-decreasing space density prior. In other cases (i.e., where distances are estimated from the spectral type of the donor), we simply adopt the reported distance and uncertainty.

Some binaries have precise proper motions measured using radio interferometry. For the remaining systems, we use proper motions from \textit{Gaia} DR3 whenever available. Some associated sources in \textit{Gaia} DR3 have available proper motions that are not constraining (i.e., \texttt{pmra}$/$\texttt{pmra\_error} or \texttt{pmdec}$/$\texttt{pmdec\_error} $\lesssim 3$); we omit these proper motions from Table~\ref{tab:catalog} and do not include such systems in our final sample.

The initial sample we considered (Table~\ref{tab:catalog}) includes 20 low- and intermediate-mass BH X-ray binaries that were discovered as X-ray transients and later dynamically confirmed. We take this list of dynamically confirmed BH X-ray transients from BlackCAT \citep{corral-santana_blackcat_2016}. We also include the persistent X-ray source Cyg X-1, which is the only unambiguous high-mass BH X-ray binary in the Milky Way known to date. We supplement the sample with Gaia BH1, Gaia BH2, and Gaia BH3, three non-interacting wide BH-luminous star binaries discovered using \textit{Gaia} astrometry. We additionally include recent BH candidates in the Milky Way (HD 130298) and LMC (VFTS 243) that are argued to contain a massive unseen compact object on dynamical grounds, as well as a spectroscopic BH candidate discovered in a wide binary in the globular cluster NGC 3201.\footnote{We take the list of dormant BH candidates from the catalog maintained at \texttt{https://mkenne15.github.io/BHCAT/index.html}.} Finally, we include LMC X-1 and LMC X-3, persistent X-ray sources that are the only known high-mass BH X-ray binaries in the LMC. The full catalog is provided in Table~\ref{tab:catalog}.

\begin{deluxetable*}{cccccccccc}
\tablecaption{Catalog of literature proper motions, distances, and center-of-mass radial velocities for dynamically confirmed BHs with luminous companions (LCs) in the Milky Way and Large Magellanic Cloud (LMC). The distance to BH-LC systems in the LMC is constrained by eclipsing binaries (EBs). References for literature values are provided as superscripts. We only include objects for which an analysis based on the properties of the local stellar population is feasible in our final sample (see Table \ref{tab:kicks}). These objects are identified in bold. \label{tab:catalog}}
\tablehead{\colhead{Name} & \colhead{$\alpha$ (h:m:s)} & \colhead{$\delta$ ($^{\circ}$:':'')} & \colhead{$\mu_{\alpha} \cos{\delta}$ (mas yr$^{-1}$)} & \colhead{$\mu_{\delta}$ (mas yr$^{-1}$)} & \colhead{$d$ (kpc)} & \colhead{Method} & \colhead{$\gamma$ (km s$^{-1}$)} \\
\colhead{(1)} & \colhead{(2)} & \colhead{(3)} & \colhead{(4)} & \colhead{(5)} & \colhead{(6)} & \colhead{(7)} & \colhead{(8)}}
\startdata
\textbf{Swift J1727.8-162} & 17:27:43.31 & -16:12:19.23 & $-0.04 \pm 0.60$$^1$ & $-4.92 \pm 0.43$$^1$ & $3.7 \pm 0.3$$^2$ & Various & $-178 \pm 3$$^2$ \\
\textbf{MAXI J1820+070} & 18:20:21.94 & +07:11:07.19 & $-3.051 \pm 0.046$$^3$ & $-6.394 \pm 0.075$$^3$ & $2.96 \pm 0.33$$^3$ & Radio Parallax & $-21.6 \pm 2.3$$^4$ \\
\textbf{MAXI J1305-704} & 13:06:55.30  & -70:27:05.11 & $-7.89 \pm 0.62$$^1$ & $-0.16 \pm 0.72$$^1$ & $7.5^{+1.8^5}_{-1.4}$ & Spectroscopic & $-9 \pm 5$$^5$ \\
XTE J1650-500 & 16:50:00.98 & -49:57:43.60 & --- & --- & $2.6 \pm 0.7$$^6$ & X-ray & $19 \pm 3$$^7$ \\
\textbf{XTE J1118+480} & 11:18:10.79  & +48:02:12.42 & $-18.105 \pm 0.155$$^1$ & $-6.687 \pm 0.217$$^1$ & $1.72 \pm 0.10$$^8$ & Ellipsoidal & $2.7 \pm 1.1$$^9$ \\
XTE J1859+226 & 18:58:41.58 & +22:39:29.40 & --- & --- & $6.3 \pm 1.7$$^{10}$ & Ellipsoidal & $115 \pm 42$$^{11}$ \\
\textbf{V4641 Sgr} & 18:19:21.58  & -25:24:25.10  & $-0.779 \pm 0.026$$^1$ & $+0.433 \pm 0.020$$^1$ & $4.739_{-0.602}^{+0.766^{12}}$ & Optical Parallax & $107.4 \pm 2.9$$^{13}$ \\
XTE J1550-564 & 15:50:58.70  & -56:28:35.20 & --- & --- & $4.38^{+0.58^{14}}_{-0.41}$ & Ellipsoidal & $-68 \pm 19$$^{15}$ \\
\textbf{GRO J1655-40} & 16:54:00.14  & -39:50:44.90 & $-3.3 \pm 0.5$$^{16}$ & $-4.0 \pm 0.4$$^{16}$ & $3.2 \pm 0.2$$^{17}$ & Jets & $-167.1 \pm 0.6$$^{18}$ \\
GRS 1009-45 & 10:13:36.34  & -45:04:31.50 & --- & --- & $3.8 \pm 0.3$$^{19}$ & Ellipsoidal & $30.1 \pm 5.0$$^{20}$ \\
GRS 1915+105 & 19:15:11.55 & +10:56:44.80 & $-3.14 \pm 0.03$$^{21}$ & $-6.23 \pm 0.04$$^{21}$ & $9.4 \pm 1.0$$^{21}$ & Radio Parallax & $12.3 \pm 1.0$$^{22}$ \\
GRO J0422+32 & 04:21:42.79 &  +32:54:27.10 & --- & --- & $2.49 \pm 0.30$$^{23}$ & Ellipsoidal & $9.2 \pm 3.3$$^{24}$ \\
\textbf{GRS 1124-684} & 11:26:26.65 & -68:40:32.83 & $-2.933 \pm 0.244$$^{1}$ & $-1.392 \pm 0.262$$^{1}$ & $4.95_{-0.65}^{+0.69^{25}}$ & Ellipsoidal & $14.2 \pm 2.1$$^{26}$ \\
\textbf{V404 Cyg} & 20:24:03.82 & +33:52:01.90 & $-5.1775 \pm 0.0785$$^1$ & $-7.7776 \pm 0.0922$$^1$ & $2.39 \pm 0.14$$^{27}$ & Radio Parallax & $-2.0 \pm 0.4$$^{28}$ \\
GS 2000+251 & 20:02:49.48 & +25:14:11.36 & --- & --- & $2.7 \pm 0.7$$^{29}$ & Ellipsoidal & $18.9 \pm 4.2$$^{30}$ \\
GS 1354-64 & 13:58:09.70 & -64:44:05.80 & $-5.072 \pm 0.631$$^{1}$ & $-2.105 \pm 0.582$$^{1}$ & $\geq 25$$^{31}$ & Ellipsoidal & $102.0 \pm 4.0$$^{31}$ \\
H 1705-250 & 17:08:14.52 & -25:05:30.15 & --- & --- & $8.6 \pm 2.2$$^{29}$ & Ellipsoidal & $-41.1 \pm 0.8$$^{32}$ \\
\textbf{3A 0620-003} & 06:22:44.50 & -00:20:44.72 & $-0.439 \pm 0.108$$^{1}$ & $-5.138 \pm 0.096$$^{1}$ & $1.487_{-0.226}^{+0.256^{12}}$ & Optical Parallax & $8.5 \pm 1.8$$^{33}$ \\
1H 1659-487 & 17:02:49.40 & -48:47:23.40 & $-3.95 \pm 0.07$$^{34}$ & $-4.71 \pm 0.06$$^{34}$ & $\geq 5$$^{35}$ & Spectroscopic & $26 \pm 2$$^{35}$ \\
4U 1543-475 & 15:47:08.32 &  -47:40:10.80 & $-7.543 \pm 0.053$$^{1}$ & $-5.356 \pm 0.046$$^{1}$ & $7.5 \pm 0.5$$^{36}$ & Ellipsoidal & $-87 \pm 3$$^{37}$ \\
\textbf{Cyg X-1} & 19:58:21.68 & +35:12:05.78 & $-3.812 \pm 0.015$$^{1}$ & $-6.31 \pm 0.017$$^{1}$ & $2.147_{-0.054}^{+0.064^{12}}$ & Optical Parallax & $-5.1 \pm 0.5$$^{38}$ \\
\textbf{Gaia BH1} & 17:28:41.09 & -00:34:51.93 & $-7.70 \pm 0.02$$^{1}$ & $-25.85 \pm 0.03$$^{1}$ & $0.477 \pm 0.004$$^{39}$ & Optical Parallax & $48.379 \pm 0.001$$^{40}$ \\
\textbf{Gaia BH2} & 13:50:16.73 & -59:14:20.42 & $-10.48 \pm 0.10$$^{1}$ & $-4.61 \pm 0.06$$^{1}$ & $1.16 \pm 0.02$$^{41}$ & Optical Parallax & $-4.22 \pm 0.13$$^{41}$ \\
Gaia BH3 & 19:39:18.71 & +14:55:54.01 & $-28.317 \pm 0.067$$^{1}$ & $-155.221 \pm 0.111$$^{1}$ & $0.590 \pm 0.006$$^{42}$ & Parallax & $-357.31 \pm 0.44$$^{42}$ \\
HD 130298 & 14:49:33.77 & -56:25:38.47 & $-6.495 \pm 0.015$$^{1}$ & $-1.200 \pm 0.016$$^{1}$ & $2.406_{-0.078}^{+0.085^{12}}$ & Optical Parallax & $-36.54 \pm 0.36$$^{43}$ \\
NGC 3201 \#12560 & 10:17:37.09 & -46:24:55.33 & --- & --- & $4.55 \pm 0.20$$^{44}$ & Optical Parallax & $494.5 \pm 2.4$$^{45}$ \\
VFTS 243 & 05:38:08.41 & -69:09:18.98 & $+1.722 \pm 0.034$$^{1}$ & $+0.603 \pm 0.032$$^{1}$ & $50 \pm 1$$^{46}$ & EBs & $260.2 \pm 0.9$$^{47}$ \\
LMC X-1 & 05:39:38.70 & -69:44:36.00 & $+1.889 \pm 0.020$$^{1}$ & $+0.622 \pm 0.023$$^{1}$ & $50 \pm 1$$^{46}$ & EBs & $21.0 \pm 4.8$$^{48}$ \\
LMC X-3 & 05:38:56.63 & -64:05:03.32 & $+1.678 \pm 0.065$$^{1}$ & $+0.449 \pm 0.066$$^{1}$ & $50 \pm 1$$^{46}$ & EBs & $303 \pm 2$$^{49}$ \\
\enddata
\tablerefs{[1] \citet{gaia_collaboration_gaia_2023}, [2] \citet{matasanchez_2024}, [3] \citet{atri_2020}, [4] \citet{torres_2019}, [5] \citet{matasanchez_2021}, [6] \citet{homan_2006}, [7] \citet{sanchezfernandez_2002}, [8] \citet{gelino_2006}, [9] \citet{gonzalezhernandez_xte_2008}, [10] \citet{hynes_2002}, [11] \citet{yanesrizo_2022}, [12] \citet{bailerjones_2021}, [13] \citet{orosz_2001}, [14] \citet{orosz_2011}, [15] \citet{orosz_2002}, [16] \citet{mirabel_2002}, [17] \citet{hjellming_1995}, [18] \citet{gonzalezhernandez_gro_2008}, [19] \citet{gelino_2002}, [20] \citet{filippenko_1999}, [21] \citet{reid_2023}, [22] \citet{reid_2014}, [23] \citet{gelino_2003}, [24] \citet{filippenko_1995}, [25] \citet{wu_2016}, [26] \citet{wu_2015}, [27] \citet{millerjones_first_2009}, [28] \citet{casares_2019}, [29] \citet{barret_1996}, [30] \citet{harlaftis_1996}, [31] \citet{casares_2009}, [32]  \citet{filippenko_1997}, [33] \citet{gonzalezhernandez_2010}, [34] \citet{atri_potential_2019}, [35] \citet{heida_2017}, [36] \citet{orosz_4U_2002}, [37] \citet{orosz_1998}, [38] \citet{gies_2008}, [39] \citet{el-badry_sun-like_2023}, [40] \citet{nagarajan_espresso_2024}, [41] \citet{el-badry_red_2023}, [42] \citet{gaia_collaboration_discovery_2024}, [43] \citet{mahy_identifying_2022}, [44] \citet{vasiliev_2021}, [45] \citet{giesers_detached_2018}, [46] \citet{pietrzynski_2013}, [47] \citet{shenar_x-ray-quiet_2022}, [48] \citet{hyde_2017}, [49] \citet{orosz_2014}}
\end{deluxetable*}

\begin{figure*}
\epsscale{1.0}
\plotone{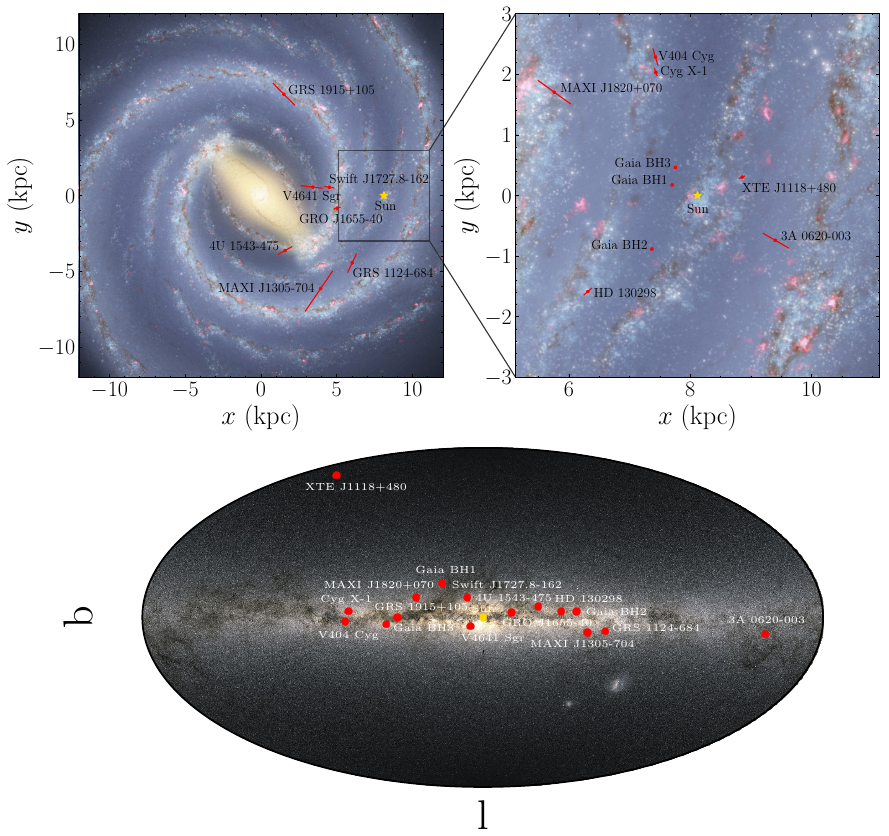}
\caption{Top: Distribution of binaries hosting dynamically confirmed BHs with luminous companions in a face-on artist's rendition of the Milky Way disk  (Credit: NASA/JPL-Caltech/ESO/R. Hurt). The right panel shows an inset zoomed in on the solar neighborhood. Error bars correspond to uncertainties on distance measurements. Systems without reliable proper motions or distances are not included, resulting in most BHs in the Galactic bulge being omitted from the figure. Bottom: Sky map of the binaries shown in the top panels. The scatter points representing the locations of the targets in Galactic coordinates are superimposed on a background map of the Milky Way from \textit{Gaia} DR3.}
\label{fig:mw_distribution}
\end{figure*}

We visualize the Galactic distribution of BH-LC systems with reliable proper motions and distances in Figure~\ref{fig:mw_distribution}. In the top panels, we use a face-on artist's rendition of the Milky Way disk and show error bars derived from uncertainties on distance estimates from the literature. In the bottom panel, we show a sky map of these same systems, superimposed on a background map of the Milky Way from \textit{Gaia} DR3. Since systems without reliable proper motion or distance estimates are not considered, the vast majority of sources included in Figure~\ref{fig:mw_distribution} are in the disk rather than the Galactic bulge.

In our final sample, we remove objects for which a kinematic analysis based on the properties of the local stellar population is infeasible. These include 4U 1543-475, which has a literature distance that places it in the bulge; GRS 1915+105, which is behind too much dust to allow for a robust local comparison sample from \textit{Gaia} DR3; Gaia BH3, which is a halo source associated with the ED-2 stellar stream \citep{balbinot_2024}; and systems in globular clusters or the LMC. We also do not consider HD 130298, which has not yet been conclusively proven to host a stellar-mass BH. After filtering out sources with missing values, we are left with 12 systems, 7 of which have geometric distance measurements and 5 of which have distance measurements based on modeling of the donor. Our final sample size is limited by a lack of well-constrained distance and proper motion measurements in the literature. The final sample is provided in Table~\ref{tab:kicks}.

\section{Analysis} \label{sec:analysis}

\subsection{Constructing the Reference Samples}
\label{sec:construction}

For each of our targets, we use \textit{Gaia} DR3 to construct a comparison sample from the local stellar population. Here, ``local'' is defined as being both within a projected radius of 500 pc and within a distance range of $\left(D - \Delta D, D + \Delta D\right)$, where $D \pm \Delta D$ represents the distance measurement from the literature. We construct this sample by retrieving sources from the \textit{Gaia} DR3 catalog that fall within these limits and have published parallaxes, proper motions, and radial velocities. Rather than simply inverting the parallax to determine the distance to each source, we use the geometric distances derived by \citet{bailerjones_2021} based on direction-dependent priors. To ensure a high quality comparison sample, we require that \texttt{parallax\_over\_error} $\geq 3$, radial velocity error $< 5$ km s$^{-1}$, and tangential velocity error $< 30$ km s$^{-1}$ for the sources in our final comparison sample. Here, the tangential velocity error is calculated from the astrometric uncertainties and correlation coefficients reported in \textit{Gaia} DR3. We adopt a cutoff of $30$ km s$^{-1}$ on the uncertainty in the tangential velocity to ensure that it is not greater than the velocity dispersion in the Galactic disk \citep[see e.g.,][]{viera_2022}. Most of the stars in the comparison sample are red giants, since those are the sources bright enough to have RVs measured by \textit{Gaia}. An implicit assumption in our analysis is that the BH binary and the reference samples probe similar stellar populations. This assumption may break down; e.g., if a BH binary is born in the disk and is subsequently kicked into the halo. We discuss this caveat further in Section~\ref{sec:caveats}.

We use Astropy \citep{2013A&A...558A..33A, 2018AJ....156..123A} to convert the space velocities of the target and each member of the local comparison sample into cylindrical coordinates in a Galactocentric reference frame. We assume a solar space velocity of $\left(V_{R, \odot}, V_{\phi, \odot}, V_{z, \odot}\right) = (-12.9, 245.6, 7.78)$ km s$^{-1}$ \citep{drimmel_poggio_2018}. We also adopt a distance to the Galactic center of $R_0 = 8.122$ kpc \citep{gravity_2018} and a solar height above the galactic plane of $z_{\odot} = 20.8$ pc \citep{bennett_jo_2019}. To account for (potentially asymmetric) measurement errors, we draw 10,000 Monte Carlo samples from split normal distributions $\mathcal{SN}(\mu, \sigma_1, \sigma_2)$ for the distance, proper motions, and center-of-mass radial velocity, respectively, using uncertainties reported in the literature. The split normal distribution, which joins two normal distributions with the same mode $\mu$ but different variances $\sigma_1^2$ and $\sigma_2^2$, has a probability density function given by:

\begin{eqnarray}
    f(x; \mu, \sigma_1, \sigma_2) = 
       \sqrt{\frac{2}{\pi (\sigma_1 + \sigma_2)^2}} \nonumber \\ \cases{
    \exp{\left(-\frac{(x - \mu)^2}{2 \sigma_1^2}\right)} & for $x < \mu$ \cr
    \exp{\left(-\frac{(x - \mu)^2}{2 \sigma_2^2}\right)} & otherwise. \cr
    }
\end{eqnarray}

We use these samples to determine the 16th, 50th, and 84th percentiles of the space velocities in cylindrical coordinates. We use the median velocities as our fiducial values throughout our analysis.

For the purpose of comparison with previous work, we compute the local present-day peculiar velocity of each system as follows:

\begin{equation}
\label{eq:local_pec}
    V_{\text{pec, local}} = \sqrt{V_R^2 + V_z^2 + \left(V_{\phi} - \tilde{V}_{\phi}\right)^2}
\end{equation}

where $\left(V_R, V_z, V_{\phi}\right)$ is the space velocity of the system expressed in cylindrical coordinates in a Galactocentric reference frame, and $\tilde{V}_{\phi}$ is the median azimuthal velocity of the local comparison sample. 

Assuming the Milky Way gravitational potential described by \texttt{MilkyWayPotential2022} in \texttt{gala} \citep{gala}, we also compute the present-day peculiar velocity for each system relative to the circular velocity at the corresponding Galactocentric radius. This circular velocity, denoted as $V_{\text{circ}}\left(R, z = 0\right)$, is evaluated at the Galactic plane, rather than at the actual location of each system:

\begin{equation}
\label{eq:circ_pec}
     V_{\text{pec, circ}} = \sqrt{V_R^2 + V_z^2 +  \left[V_{\phi} - V_{\text{circ}}\left(R, z = 0\right) \right]^2}\,.
\end{equation}

The two estimates of the present-day peculiar velocity differ slightly, with $V_{\text{pec, circ}}$ being greater than, identical to, and lesser than $V_{\text{pec, local}}$ for seven, one, and four systems, respectively. We provide both of these peculiar velocities in Table~\ref{tab:kicks}, but emphasize that they do not represent our fiducial estimates of the BH natal kicks. Unless specified otherwise, we adopt $V_{\text{pec, local}}$ as our fiducial estimate of the present-day peculiar velocity.

\subsection{Accounting for the Velocity Dispersion of the Local Stellar Population}
\label{sec:dispersion}

\begin{figure*}
\epsscale{1.0}
\plotone{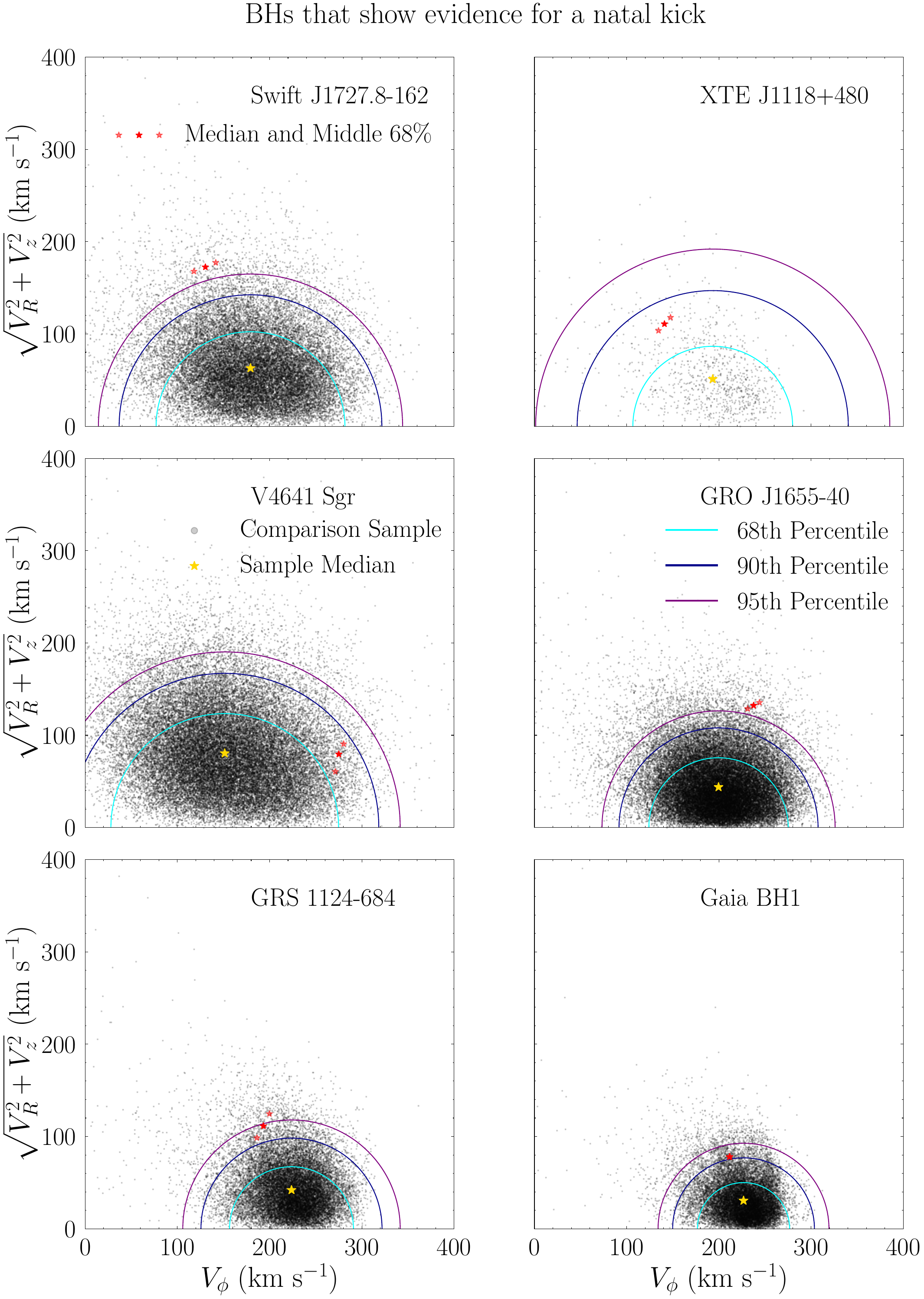}
\caption{Toomre diagrams for systems hosting BHs that show evidence for a natal kick. The azimuthal and non-azimuthal components of the spatial velocities of the targets are compared to those of a local comparison sample of stars from \textit{Gaia} DR3. The median of the comparison sample is marked with a gold star. Semi-circles centered on the median $V_{\phi}$ that contain 68\%, 90\%, and 95\% of the comparison sample are plotted in cyan, dark blue, and purple, respectively. The fiducial location of each system based on literature distances, proper motions, and radial velocities is plotted with a red star, with lower and upper bounds on the location marked by red stars of higher transparency. Each of the systems shown here lies beyond the 68\% threshold, implying with 1$\sigma$ evidence that their stellar-mass BHs received a natal kick.}
\label{fig:kick_toomre_plots}
\end{figure*}

\begin{figure*}
\epsscale{1.0}
\plotone{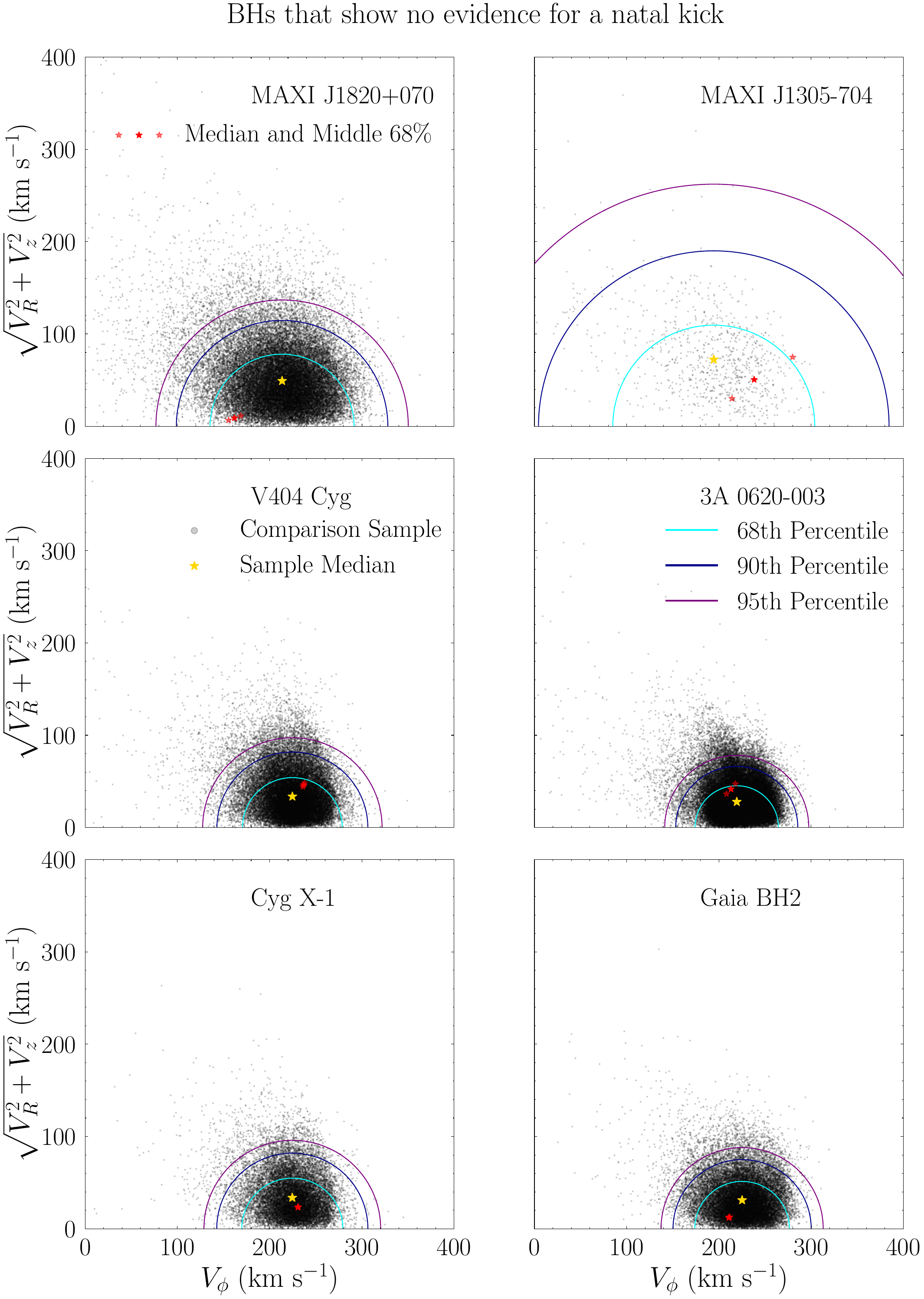}
\caption{Same as Figure \ref{fig:kick_toomre_plots}, but for systems hosting BHs that show no evidence for a natal kick. Each of the systems shown here lies below the 68\% threshold, implying that their peculiar velocities are consistent with the velocity dispersion of the local stellar population.}
\label{fig:no_kick_toomre_plots}
\end{figure*}

We construct Toomre diagrams by plotting $\sqrt{V_R^2 + V_z^2}$ versus $V_{\phi}$ for both the target and the comparison sample. We present these Toomre diagrams in Figures~\ref{fig:kick_toomre_plots} and \ref{fig:no_kick_toomre_plots}.

To account for the velocity dispersion of the local stellar population, we calculate the median azimuthal velocity $\tilde{V}_{\phi}$ of the comparison sample, and determine the radii, $V_p$, of semi-circles centered on $(\tilde{V}_{\phi}, 0)$ that include 68\%, 90\% and 95\% of sources, respectively. While any choice of percentile is arbitrary, the values we use are motivated by common choices for confidence intervals. We list these radii in Table~\ref{tab:kicks}, and overplot the corresponding semi-circles on our Toomre diagrams in Figures~\ref{fig:kick_toomre_plots} and \ref{fig:no_kick_toomre_plots}.

Choosing the 68\% threshold as a $1\sigma$ estimate of the velocity dispersion of the local stellar population, we calculate the minimum natal kick of each BH as the difference between its 3D space velocity 

\begin{equation}
    V = \sqrt{V_R^2 + V_z^2 + V_{\phi}^2}
\end{equation}

and this threshold, i.e.\

\begin{equation}
\label{eq:min_kick}
    V_{\text{kick}, \min} = \max\left(0, V - V_{68\%}\right).
\end{equation}

We present our fiducial minimum natal kicks in Table~\ref{tab:kicks}. We also provide the velocity percentile of each system, representing the fraction of stars in each comparison sample that are less kinematically hot than the BH. Half of the BH systems lie beyond the 68\% threshold on the Toomre diagram, suggesting weak evidence of a natal kick. In addition, one-third and one-sixth of systems have space velocities outside the 90th and 95th percentile of the comparison population, respectively, providing stronger evidence of a kick. The other systems lie below this threshold, implying that their spatial velocities are consistent with the local stellar population and that their kinematics are consistent with little to no natal kick. This does not rule out a kick for these systems, but it suggests that any kick was weaker than the velocity dispersion of the local comparison sample, which is typically $30$--$50$ km s$^{-1}$.

\begin{deluxetable*}{cccccccc}
\tablecaption{Local velocity dispersion and natal kick estimates for dynamically confirmed BHs with luminous companions. The columns $V_p$ represent the radii of semi-circles centered on $(\tilde{V}_{\phi}, 0)$ that include $p$\% of sources in the comparison sample, where $\tilde{V}_{\phi}$ is the sample's median azimuthal velocity (see Section \ref{sec:dispersion}). The minimum natal kick $V_{\text{kick}, \min}$ (Equation \ref{eq:min_kick}) is computed relative to $V_{68\%}$. Errors on present-day peculiar velocities $V_{\text{pec, local}}$ and $V_{\text{pec, circ}}$ (Equations \ref{eq:local_pec} and \ref{eq:circ_pec} respectively) are derived from 16th and 84th percentiles of Monte Carlo samples (see Section \ref{sec:construction}). The velocity percentile represents the fraction of stars in each comparison sample that are less kinematically hot than the BH. \label{tab:kicks}}
\tablehead{\colhead{Name} & \colhead{$V_{68\%}$} & \colhead{$V_{90\%}$} & \colhead{$V_{95\%}$} & \colhead{$V_{\text{pec, local}}$} & \colhead{$V_{\text{pec, circ}}$} & \colhead{$V_{\text{kick}, \min}$} & Velocity Percentile \\
& (km s$^{-1}$) & (km s$^{-1}$) & (km s$^{-1}$) & (km s$^{-1}$) & (km s$^{-1}$) & (km s$^{-1}$) & \\
\colhead{(1)} & \colhead{(2)} & \colhead{(3)} & \colhead{(4)} & \colhead{(5)} & \colhead{(6)} & \colhead{(7)} & \colhead{(8)}}
\startdata
Swift J1727.8-162 & $102$ & $142$ & $165$ & $180_{-5}^{+5}$ & $186_{-5}^{+6}$ & $77$ & $97$ \\
MAXI J1820+070 & $78$ & $115$ & $137$ & $53_{-7}^{+7}$ & $33_{-7}^{+7}$ & $0$ & $37$ \\
MAXI J1305-704 & $110$ & $190$ & $262$ & $71_{-28}^{+41}$ & $74_{-29}^{+42}$ & $0$ & $32$ \\
XTE J1118+480 & $87$ & $147$ & $192$ & $123_{-9}^{+9}$ & $119_{-9}^{+9}$ & $36$ & $84$ \\
V4641 Sgr & $123$ & $167$ & $190$ & $147_{-5}^{+4}$ & $105_{-9}^{+7}$& $23$ & $82$ \\
GRO J1655-40 & $76$ & $108$ & $126$ & $138_{-3}^{+2}$ & $138_{-3}^{+3}$ & $62$ & $97$ \\
GRS 1124-684 & $67$ & $98$ & $118$ & $115_{-14}^{+15}$ & $111_{-12}^{+13}$ & $48$ & $94$ \\
V404 Cyg & $54$ & $82$ & $97$ & $48_{-2}^{+2}$ & $68_{-2}^{+2}$ & $0$ & $60$ \\
3A 0620-003 & $45$ & $66$ & $78$ & $42_{-6}^{+7}$ & $53_{-2}^{+2}$ & $0$ & $62$ \\
Cyg X-1 & $54.8$ & $81.9$ & $95.8$ & $24.2_{-0.3}^{+0.4}$ & $49.3_{-0.5}^{+0.5}$ & $0$ & $18$ \\
Gaia BH1 & $50.0$ & $76.9$ & $92.6$ & $79.4_{-0.3}^{+0.3}$ & $81.56_{-0.04}^{+0.04}$ & $29$ & $91$ \\
Gaia BH2 & $51.3$ & $74.9$ & $87.9$ & $18.6_{-0.9}^{+0.9}$ & $25.7_{-0.3}^{+0.3}$ & $0$ & $10$ \\
\enddata
\end{deluxetable*}

The Toomre diagrams for systems hosting BHs that show at least weak evidence for a kick are presented in Figure~\ref{fig:kick_toomre_plots}, while the Toomre diagrams for systems hosting BHs that show no evidence for a natal kick are presented in Figure~\ref{fig:no_kick_toomre_plots}. The median of the comparison sample is marked with a gold star. The fiducial location of the target on each Toomre diagram is marked with a red star. The upper and lower bounds on $\left(V_{\phi}, \sqrt{V_R^2 + V_z^2}\right)$ (derived from the 16th and 84th percentiles, see Section~\ref{sec:construction}) are plotted with red stars of higher transparency.

This method of computing a minimum natal kick assumes that each system was born in the Galactic disk, and has been heated by both the BH natal kick and dynamical processes over the course of its lifetime. For comparison purposes, we also compute a more stringent estimate of the minimum natal kick that assumes that each system was born in its observed present-day location, and has only been heated by its natal kick. In the modified approach, we determine the radii of circles (rather than the semi-circles in our fiducial analysis) centered on $\left(\tilde{V}_{\phi}, \tilde{V}_{\perp}\right)$ that include 68\%, 90\%, and 95\% of sources, respectively, where $\tilde{V}_{\perp}$ is the median non-azimuthal velocity of the comparison sample, and we have defined $V_{\perp} \equiv \sqrt{V_R^2 + V_z^2}$ for brevity. The modified local present-day peculiar velocity is then given by:

\begin{equation}
\label{eq:mod_pec}
    V_{\text{pec, local}}' = 
    \sqrt{
    \left(V_{\perp} - \tilde{V}_{\perp}\right)^2 + \left(V_{\phi} - \tilde{V}_{\phi}\right)^2
    }\,.
\end{equation}

This quantity is essentially the velocity difference between the red and yellow stars in Figures~\ref{fig:kick_toomre_plots} and \ref{fig:no_kick_toomre_plots}. We provide these radii and peculiar velocities in Table~\ref{tab:circle_kicks}. We then use the new 68\% threshold as a $1\sigma$ estimate of the velocity dispersion of the local stellar population, and re-apply Equation~\ref{eq:min_kick} to calculate the minimum natal kicks reported in Table~\ref{tab:circle_kicks}. We also report the modified velocity percentiles of each BH system. As expected, we find that accounting for the median $V_{\perp}$ of the local stellar population tends to slightly decrease the magnitude of $V_{\text{kick}, \min}$. However, the two metrics generally lead to consistent kick versus no kick samples, except for MAXI J1820+070 (see Section \ref{sec:lit_comparison} for a discussion of this individual system).

\begin{deluxetable*}{ccccccc}
\tablecaption{Local velocity dispersion and natal kick estimates re-calculated after accounting for the median non-azimuthal velocity $\tilde{V}_{\perp}$ of the comparison sample. The columns $V'_p$ now represent the radii of circles centered on $\left(\tilde{V}_{\phi}, \tilde{V}_{\perp}\right)$ that include $p$\% of sources in the comparison sample, and the modified minimum natal kick $V'_{\text{kick}, \min}$ is computed relative to $V'_{68\%}$ (see discussion at the end of Section \ref{sec:dispersion}). Errors on the modified local present-day peculiar velocity $V'_{\text{pec, local}}$ (Equation \ref{eq:mod_pec}) are derived from 16th and 84th percentiles of Monte Carlo samples (see Section \ref{sec:construction}). As before, the velocity percentile represents the fraction of stars in each comparison sample that are less kinematically hot than the BH. \label{tab:circle_kicks}}
\tablehead{\colhead{Name} & \colhead{$V'_{68\%}$} & \colhead{$V'_{90\%}$} & \colhead{$V'_{95\%}$} & \colhead{$V'_{\text{pec, local}}$} & \colhead{$V'_{\text{kick}, \min}$} & Velocity Percentile \\
& (km s$^{-1}$) & (km s$^{-1}$) & (km s$^{-1}$) & (km s$^{-1}$) & (km s$^{-1}$) & \\
\colhead{(1)} & \colhead{(2)} & \colhead{(3)} & \colhead{(4)} & \colhead{(5)} & \colhead{(6)} & \colhead{(7)}}
\startdata
Swift J1727.8-162 & $69$ & $105$ & $124$ & $121_{-6}^{+6}$ & $51$ & $94$ \\
MAXI J1820+070 & $52$ & $82$ & $103$ & $66_{-5}^{+4}$ & $14$ & $82$ \\
MAXI J1305-704 & $69$ & $145$ & $223$ & $56_{-19}^{+33}$ & $0$ & $46$ \\
XTE J1118+480 & $60$ & $110$ & $159$ & $79_{-9}^{+10}$ & $19$ & $82$ \\
V4641 Sgr & $80$ & $118$ & $137$ & $124_{-3}^{+7}$ & $44$ & $92$ \\
GRO J1655-40 & $53$ & $80$ & $96$ & $96_{-3}^{+3}$ & $44$ & $95$ \\
GRS 1124-684 & $44$ & $70$ & $87$ & $76_{-14}^{+15}$ & $32$ & $92$ \\
V404 Cyg & $35$ & $57$ & $71$ & $18_{-2}^{+2}$ & $0$ & $26$ \\
3A 0620-003 & $31$ & $46$ & $57$ & $15_{-6}^{+7}$ & $0$ & $24$ \\
Cyg X-1 & $35.8$ & $56.9$ & $70.1$ & $12.0_{-0.3}^{+0.3}$ & $0$ & $12$ \\
Gaia BH1 & $33.4$ & $55.8$ & $69.3$ & $49.8_{-0.3}^{+0.3}$ & $16$ & $87$\\
Gaia BH2 & $34.7$ & $54.0$ & $65.2$ & $23.4_{-0.2}^{+0.2}$ & $0$ & $42$ \\
\enddata
\end{deluxetable*}

\subsection{Galactic Orbits}

\begin{figure*}
\epsscale{1.0}
\plotone{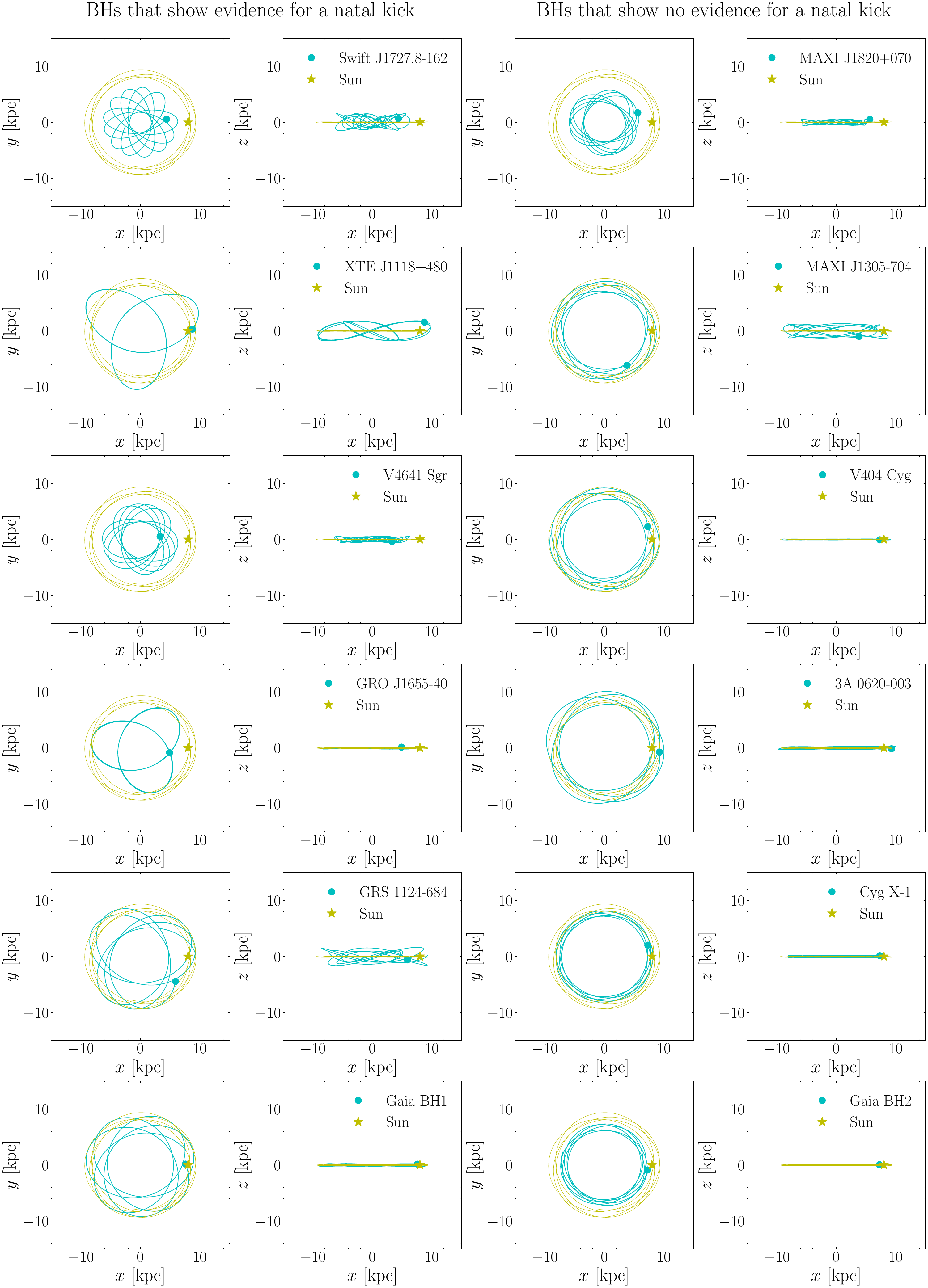}
\caption{Galactic orbits of all binaries in our final sample. The orbits are integrated back in time by 1 Gyr based on current location and spatial velocity. The solar orbit is shown for reference. In the left two columns, we show the integrated orbits for systems hosting BHs that show evidence for a natal kick. These orbits are generally puffier and less circular than the solar orbit, implying a natal kick. In the right two columns, we show the integrated orbits for systems hosting BHs that show no evidence for a natal kick. These orbits are generally close to circular and do not travel far from the disk midplane, reinforcing a lack of evidence for a natal kick.}
\label{fig:all_galactic_orbits}
\end{figure*}

Using the best-fit right ascensions, declinations, proper motions, center-of-mass RVs, and distance measurements from the literature, we compute the Galactic orbits of each binary and present them in Figure~\ref{fig:all_galactic_orbits}. We assume the Milky Way gravitational potential described by \texttt{MilkyWayPotential2022} in \texttt{gala} \citep{gala} and use $\texttt{galpy}$ \citep{bovy_2015} to integrate the Galactic orbit back in time by 1 Gyr. We compare these orbits to the solar orbit. 

\begin{deluxetable*}{ccccc}[t!]
\tablecaption{Summary of natal kick diagnostics for all BHs in our sample. $V_{\text{kick}, \min}$ is the minimum natal kick estimate from Table \ref{tab:kicks}, while $V'_{\text{kick}, \min}$ is the minimum natal kick estimate from Table \ref{tab:circle_kicks}. The former assumes that the system was born in the Galactic disk and has been heated by both the natal kick and dynamical processes, while the latter assumes that the system was born in its present-day location and has only been heated by its natal kick. We also consider the visual properties of the Galactic orbits in Figure \ref{fig:all_galactic_orbits}, focusing on the eccentricity relative to the solar orbit in the $x$-$y$ plane and the puffiness relative to the solar orbit in the $x$-$z$ plane. \label{tab:proxies}}
\tablehead{\colhead{Name} & \colhead{$V_{\text{kick}, \min}$ (km s$^{-1}$)} &  \colhead{$V'_{\text{kick}, \min}$ (km s$^{-1}$)} & Galactic Orbit in $x$-$y$ Plane & Galactic Orbit in $x$-$z$ Plane \\
\colhead{(1)} & \colhead{(2)} & \colhead{(3)} & \colhead{(4)} & \colhead{(5)}}
\startdata
Swift J1727.8-162 & \checkmark & \checkmark & \checkmark & \checkmark \\
MAXI J1820+070 & $\times$ & \checkmark & \checkmark & \checkmark \\
MAXI J1305-704 & $\times$ & $\times$ & $\times$ & \checkmark \\
XTE J1118+480 & \checkmark & \checkmark & \checkmark & \checkmark \\
V4641 Sgr & \checkmark & \checkmark & \checkmark & \checkmark \\
GRO J1655-40  & \checkmark & \checkmark & \checkmark & $\times$ \\
GRS 1124-684  & \checkmark & \checkmark & \checkmark & \checkmark \\
V404 Cyg & $\times$ & $\times$ & $\times$ & $\times$ \\
3A 0620-003 & $\times$ & $\times$ & $\times$ & $\times$ \\
Cyg X-1 & $\times$ & $\times$ & $\times$ & $\times$ \\
Gaia BH1 & \checkmark & \checkmark & \checkmark & $\times$ \\
Gaia BH2 & $\times$ & $\times$ & $\times$ & $\times$ \\
\enddata
\end{deluxetable*}

The Galactic orbits in the left two columns of  Figure~\ref{fig:all_galactic_orbits} correspond to the systems that display evidence for a natal kick based on Figure~\ref{fig:kick_toomre_plots}. In general, these integrated orbits are puffier and more eccentric than the solar orbit, consistent with a BH kick. On the other hand, the Galactic orbits in the right two columns of Figure~\ref{fig:all_galactic_orbits} correspond to the systems that show no evidence for a natal kick based on Figure~\ref{fig:no_kick_toomre_plots}. Most of these integrated orbits are close to circular and do not travel far from the disk midplane, reinforcing a lack of evidence for a BH kick. 

However, the Galactic orbit of MAXI J1820+070 is more eccentric than the solar orbit, while the Galactic orbit of MAXI J1305-704 is puffier than that of the solar orbit. To explain this apparent tension, we point out that the shape of the Galactic orbit of a system is more indicative of its peculiar velocity than it is of its true natal kick (i.e., after accounting for the local velocity dispersion). Furthermore, in cases where measurement uncertainties are large, assuming a different fiducial value for the proper motion, systemic velocity, or distance could significantly alter the shape of the integrated orbit. We emphasize that we are not claiming that MAXI J1820+070 or MAXI J1305-704 definitely formed without a kick, but rather that their kinematics do not provide strong evidence for a kick. We provide a summary of the different natal kick indicators for each BH system in Table \ref{tab:proxies}.

\section{Discussion} \label{sec:discussion}

\subsection{Interpretation of Results}
\label{sec:interpretation}

If we use $V_{68\%}$ as our threshold, we find that six of the systems in our final sample  (Swift J1727.8-162, XTE J1118+480, V4641 Sgr, GRO J1655-40, GRS 1124-684, and Gaia BH1) show evidence for a natal kick, while the other six (MAXI J1820+070, MAXI J1305-704, V404 Cyg, 3A 0620-003, Cyg X-1, and Gaia BH2) do not. Four of these systems (Swift J1727.8-162, GRO J1655-40, GRS 1124-684, and Gaia BH1) show evidence of a natal kick if $V_{90\%}$ is applied as a more stringent threshold, and two of these systems (Swift J1727.8-162 and GRO J1655-40) show evidence of a natal kick even if $V_{95\%}$ is used a threshold instead. 

Due to sample selection effects, it would be a stretch to conclude from these results that $50\%$ of stellar-mass BHs in binaries were born with a kick. However, it is clear that at least some BHs were likely born with a kick. To illustrate, the binomial probabilities of observing six systems with a higher peculiar velocity than 68\% of the comparison sample, four systems with a higher peculiar velocity than 90\% of the comparison sample, and two systems with a higher peculiar velocity than 95\% of the comparison sample are only about $0.098$, $0.021$, and $0.099$, respectively.

It is also clear that at least some BHs were born without a kick. Specifically, V404 Cyg and VFTS 243 (which is not one of the objects in our kinematic sample) must have been born with very weak kicks. Their unique orbits allow for tighter kick constraints than can be achieved from kinematics alone. V404 Cyg was recently reported to be part of a hierarchical triple, and only a natal kick $\lesssim 5$ km s$^{-1}$ allows the system to remain bound \citep{burdge_2024, shariat_2024}. Since VFTS 243 has not yet been tidally synchronized, its near-circular orbit implies a natal kick $\lesssim 10$ km s$^{-1}$ \citep{Stevance2023, vigna-gomez_constraints_2024}.

\subsection{Comparison to Studies of Individual Systems}
\label{sec:lit_comparison}

Several studies in the literature have investigated the natal kicks of individual BH systems.  We compare our results against those of these studies below. In general, we find that our dichotomous classification of the BH systems considered in this work is in broad agreement with the literature. However, based on our computed Toomre diagrams, we find that some systems that have been identified in the literature as requiring a natal kick to explain their peculiar velocities do not actually need one at all --- see the discussion of V404 Cyg for a compelling example. The minimum kicks we infer for most systems are also weaker than their corresponding peculiar velocities, which have often been interpreted as kick velocities in previous work.

\subsubsection{Swift J1727.8-162}

\citet{matasanchez_2024} use the Galactocentric potential \texttt{MWPotential2014} from \texttt{galpy} \citep{bovy_2015} to calculate a present-day peculiar velocity for Swift J1727.8-162 of $207 \pm 7$ km s$^{-1}$. This peculiar velocity is computed relative to $V_{\text{circ}}(R, z = 0)$, or the rotational velocity evaluated at the point corresponding to the projection of the actual location of Swift J1727.8-162 down onto the Galactic plane.

Since our comparison sample lies above the Galactic plane, and the space velocity of stars is less ordered at higher $z$, we find a lower local rotational velocity than \citet{matasanchez_2024}. This translates to a smaller local present-day peculiar velocity of $V_{\text{pec, local}} = 180 \pm 5$ km s$^{-1}$. If we use \texttt{MilkyWayPotential2022} from \texttt{gala} \citep{gala} to compute a present-day peculiar velocity for the system relative to $V_{\text{circ}}(R, z = 0)$ instead, we derive $V_{\text{pec, circ}} = 186^{+6}_{-5}$ km s$^{-1}$. The small discrepancy in the results can be attributed to the difference in assumption of Galactocentric potential. Nevertheless, \citet{matasanchez_2024} suggest that a large kick was imparted to the system at birth, in agreement with our findings.

\subsubsection{MAXI J1820+070}

\citet{atri_potential_2019} define the ``potential kick velocity'' of a system as its peculiar velocity (i.e., relative to the local standard of rest) when it crosses the Galactic plane. To compute the probability distribution of this quantity, they sample the system's present-day parameters from Gaussian distributions representing literature measurements and their associated uncertainties. Then, they integrate its Galactic orbit back in time for 10 Gyr, recording the peculiar velocity at each plane crossing. \citet{atri_2020} use the method of \citet{atri_potential_2019} to calculate a median potential kick velocity (i.e., peculiar velocity at birth) of $120$ km s$^{-1}$ for MAXI J1820+070, suggesting that the system likely received a large natal kick. In contrast, we find a much lower present-day peculiar velocity of $53 \pm 7$ km s$^{-1}$. The location of the system on a Toomre diagram is consistent with the velocity dispersion of the local stellar population.

We note that, if we use circles centered on $(\tilde{V}_{\phi}, \tilde{V}_{\perp})$ rather than semi-circles centered on $(\tilde{V}_{\phi}, 0)$ to calculate the 68\% threshold, we find that MAXI J1820+070 potentially received a small natal kick (i.e., a minimum natal kick of $> 14$ km s$^{-1}$, with a present-day peculiar velocity of $66^{+4}_{-5}$ km s$^{-1}$). Nevertheless, the orbit of MAXI J1820+070 does not constitute strong evidence for a kick, because about 20\% of stars in the same region of the Galaxy are as or even more kinematically hot. This does not rule out the possibility that MAXI J1820+070 formed with a kick, but the data are not sufficient to demonstrate that it did.

\citet{poutanen_2022} use optical polarimetry to determine a lower limit of $40^{\circ}$ on the BH spin-orbit misalignment of MAXI J1820+070. While BH natal kicks have been proposed as an explanation for observed misalignments in BH X-ray binaries, \citet{poutanen_2022} do not attempt to explain the origin of the misalignment in this system with binary evolution modeling.

\subsubsection{MAXI J1305-704}

\citet{kimball_black_2023} compute a peculiar velocity at birth for MAXI J1305-704 of $74^{+19}_{-11}$ km s$^{-1}$. \citet{matasanchez_2021} compute a present-day peculiar velocity of $80^{+30}_{-30}$ km s$^{-1}$. Both studies suggest that the kinematics of MAXI J1305-704 imply that it received a kick at birth.

Our present-day peculiar velocity of $71^{+41}_{-28}$ km s$^{-1}$ is in agreement with the result of \citet{matasanchez_2021}. However, we find that the space velocity of the system is consistent with the velocity dispersion of the local stellar population, indicating insufficient evidence for a natal kick.

\subsubsection{XTE J1118+480}

\citet{mirabel_high-velocity_2001} compute a space velocity relative to the local standard of rest of $145$ km s$^{-1}$ for XTE J1118+480, identifying it as a high-velocity BH X-ray binary. They suggest that the system may have been kicked out of a globular cluster. \citet{gualandris_has_2005} trace the system's Galactic orbit back to the plane and infer a peculiar velocity at birth of $183 \pm 31$ km s$^{-1}$, arguing for an asymmetric kick. They study the evolutionary state of the companion star, placing an upper limit on the age of the system of $2$--$5.5$ Gyr and ruling out a globular cluster origin. \citet{fragos_understanding_2009} reconstruct an evolutionary history of the system and place a lower limit of $80$ km s$^{-1}$ on the natal kick. 

We find a present-day peculiar velocity of $123 \pm 9$ km s$^{-1}$. Using $V_{68\%}$ as a representation of the velocity dispersion of the local stellar population, we estimate a minimum natal kick of $36$ km s$^{-1}$. The data are not inconsistent with a kick as large as $180$ km s$^{-1}$, but the orbit does not require one.

\subsubsection{V4641 Sgr}

\citet{salvesen_pokawanvit_2020} try to explain the extreme spin-orbit misalignment of V4641 Sgr with a natal kick model. Using the method of \citet{atri_potential_2019}, they compute a median potential kick velocity of $123$ km s$^{-1}$. Nevertheless, they find that the natal kick magnitude cannot explain the observed misalignment, and suggest instead that the jet axis does not reliably trace the black hole spin axis.

We find a present-day peculiar velocity of $147^{+4}_{-5}$ km s$^{-1}$, implying a minimum natal kick of $23$ km s$^{-1}$ at the $1\sigma$ level. While V4641 Sgr likely requires a natal kick to explain its kinematics, the magnitude of this kick is overestimated unless the local velocity dispersion is taken into account.

\subsubsection{GRO J1655-40}

GRO J1655-40 was among the first systems to be identified as a high-velocity BH X-ray binary, with \citet{brandt_effects_1995} proposing a delayed BH formation scenario to allow for an asymmetric kick that would normally be associated with a NS, and \citet{nelemans_constraints_1999} arguing for a Blauuw kick with a large mass-loss fraction instead. \citet{mirabel_2002} compute a ``runaway'' space velocity of $112 \pm 18$ km s$^{-1}$ relative to the Galactic rotation, and \citet{willems_understanding_2005} reconstruct the full evolutionary history of the binary, finding a minimum required natal kick of $\sim 30$--$50$ km s$^{-1}$.

We compute that the eccentric Galactic orbit of GRO J1655-40 implies a present-day peculiar velocity of $138^{+2}_{-3}$ km s$^{-1}$, in good agreement with \citet{mirabel_2002}. Accounting for $68\%$ of the the velocity dispersion of the local stellar population translates to a high minimum natal kick of $62$ km s$^{-1}$, second only to Swift J1727.8-162 in our sample. The value of the minimum natal kick decreases if more stringent thresholds are applied, falling to $12$ km s$^{-1}$ if $V_{95\%}$ is used instead.

\subsubsection{GRS 1124-684}

There are no dedicated studies of the natal kick of GRS 1124-684 in the literature. However, using a probabilistic estimate for the distance based on an exponential prior \citep[see e.g.,][]{gandhi_gaia_2019}, \citet{zhao_evidence_2023} find a present-day peculiar velocity of $118.6^{+15.5}_{-15.2}$ km s$^{-1}$. They also find a peculiar velocity at birth of $115.1^{+19.3}_{-18.5}$ km s$^{-1}$.

Our calculated present-day peculiar velocity of $115^{+15}_{-14}$ km s$^{-1}$ is in excellent agreement with these results. However, we emphasize that using this value as a proxy for the natal kick may overestimate the kick magnitude. Instead, accounting for the local velocity dispersion reduces the minimum natal kick for this system to $48$ km s$^{-1}$.

\subsubsection{V404 Cyg}

\citet{miller-jones_formation_2009} use very long-baseline radio interferometry to measure the proper motion of V404 Cyg, and compute a present-day peculiar velocity of $64.1 \pm 3.7^{+37.8}_{-16.6}$ km s$^{-1}$, where the error bars account for the statistical and distance uncertainties, respectively. Using models of the stellar velocity field within 2 kpc by \citet{mignard_2000}, they find a low probability of the peculiar motion being caused purely by the Galactic velocity dispersion. Based on this result, they suggest that this system received an asymmetric natal kick due to the supernova of the BH progenitor.

We compute a present-day peculiar velocity of $48 \pm 2$ km s$^{-1}$, which lies within the error bounds of \citet{miller-jones_formation_2009} and is in agreement with the results of \citet{atri_potential_2019} and \citet{zhao_evidence_2023}. In contrast to \citet{miller-jones_formation_2009}, we find that the space velocity of V404 Cyg is entirely consistent with that of the local stellar population, suggesting little to no natal kick. Indeed, \citet{burdge_2024} found that V404 Cyg is actually part of a wide hierarchical triple, and constrain any natal kick to be $\lesssim 5$ km s$^{-1}$ in order to allow the triple to remain bound. They favor a failed supernova scenario in which the BH progenitor underwent a near-complete implosion with negligible mass loss.

\subsubsection{3A 0620-003}

There are no dedicated studies of the natal kick of 3A 0620-003 in the literature. However, \citet{zhao_evidence_2023} find peculiar velocities of $43.7^{+9.9}_{-7.1}$ km s$^{-1}$ at present day and $41.8^{+11.2}_{-8.6}$ km s$^{-1}$ at birth, respectively.

Our calculated present-day peculiar velocity of $42_{-6}^{+7}$ km s$^{-1}$ is once again in excellent agreement with these results. After accounting for the velocity dispersion of the local stellar population, there is no evidence for a natal kick.

\subsubsection{Cyg X-1}

\citet{nelemans_constraints_1999} tabulate a peculiar velocity of $49 \pm 14$ km s$^{-1}$ for Cyg X-1, and suggest that a Blauuw kick due to a symmetric supernova explosion with a large mass loss fraction would be sufficient to explain its kinematic properties. \citet{wong_understanding_2012} use a stellar evolution code to reconstruct the system's evolutionary history, deriving a peculiar velocity at birth of $22$--$32$ km s$^{-1}$. They constrain the natal kick to $\lesssim 75$ km s$^{-1}$ with 95\% confidence. Comparisons of the proper motion of Cyg X-1 relative to its parent stellar association Cygnus OB3 have also led to the conclusion that it formed with little to no natal kick \citep{mirabel_cyg_2003, miller-jones_cyg_2021}. 

Our computed present-day peculiar velocity of $24.2^{+0.4}_{-0.3}$ km s$^{-1}$ is in line with these results. We find that the space velocity of Cyg X-1 is consistent with the velocity dispersion of the local stellar population, indicating a lack of evidence for a substantial natal kick. 

Most of the stars in the comparison sample we use are likely older than Cyg X-1. While an ideal comparison sample would consist of young stars of similar age, O-type stars generally do not have robust RV measurements in \textit{Gaia} DR3. In the absence of further information, the present-day peculiar velocity could be considered to be a reasonable estimate of the magnitude of Cyg X-1's natal kick.

\subsubsection{Gaia BH1}

\citet{el-badry_sun-like_2023} use a binary population synthesis code to derive a best-fit kick velocity of $16.3^{+7.1}_{-5.0}$ km s$^{-1}$ for Gaia BH1 under the implicit assumption that the system formed through isolated binary evolution. \citet{kotko_2024} use population synthesis models to investigate formation channels of Gaia BH1. For the isolated binary evolution channel, they find that assuming the natal kick velocity distribution of \citet{zhao_evidence_2023} leads to a median kick velocity of $39.3$ km s$^{-1}$.

Using $V_{68\%}$ as a representation of the velocity dispersion of the local stellar population, we estimate a minimum natal kick of $29$ km s$^{-1}$ for Gaia BH1, which falls in-between these values. \citet{el-badry_sun-like_2023} point out that the system's moderate eccentricity and wide thin disk orbit rule out large kick magnitudes (i.e., $\gtrsim 50$ km s$^{-1}$).

\subsubsection{Gaia BH2}

As with Gaia BH1, \citet{el-badry_red_2023} use a binary population synthesis code to derive natal kick constraints for Gaia BH2, finding a best-fit kick velocity of $36^{+21}_{-11}$ km s$^{-1}$. \citet{kotko_2024} also use population synthesis models to investigate formation channels of Gaia BH2. For the isolated binary evolution channel, they find that assuming the natal kick velocity distribution of \citet{zhao_evidence_2023} leads to a median kick velocity of $18.7$ km s$^{-1}$.

We find that the space velocity of Gaia BH2 is consistent with the velocity dispersion of the local stellar population, ruling out a large natal kick. Indeed, \citet{el-badry_red_2023} point out that the system's moderate eccentricity and wide thin disk orbit provide evidence against a large natal kick.

\subsection{Comparison to Other Studies of BH Natal Kicks}

In other works in the literature, the BH natal kick is computed as the peculiar velocity of the system when it crosses the Galactic plane \citep[e.g.,][]{atri_potential_2019, zhao_evidence_2023}. This approach is likely to overestimate the kick velocity, as it does not take into account the velocity dispersion of the local stellar population. We improve on these analyses by computing Toomre diagrams and accounting for this local velocity dispersion, which tends to reduce the magnitude of the inferred natal kick.

Several studies in the literature find a unimodal distribution of BH kicks \citep[e.g.,][]{atri_potential_2019, zhao_evidence_2023}. We have argued that at least a couple of BHs (V404 Cyg and VFTS 243) did not receive a natal kick, and that at least a couple of BHs show strong evidence for a natal kick (Swift J1727.8-162 and GRO J1655-40). These findings are suggestive of a bimodal kick distribution, but the uncertainties on most of the kicks are too large to rule out a broad, unimodal distribution.

Based on recent studies of V404 Cyg \citep{burdge_2024} and VFTS 243 \citep{vigna-gomez_constraints_2024}, several papers in the last year have assumed that BHs are simply born with no kick \citep[see e.g.,][]{wang_2024, stegmann_2024}. Our results imply that this is not a good assumption for all BHs: there is good evidence that at least some BHs were born with a substantial kick, with Swift J1727.8-162 and GRO J1655-40 having peculiar velocities beyond the 95th percentile of their local comparison samples from \textit{Gaia} DR3.

Almost all known BHs have masses near $\sim 10\,M_{\odot}$, with Cyg X-1 being the only system in our final sample with a BH mass robustly distinguishable from this value ($M_{\text{BH}} = 21.1 \pm 2.2\,M_{\odot}$, \citealt{miller-jones_cyg_2021}). Furthermore, BH mass measurements tend to have large error bars, as the Gaia BHs are the only systems with tight inclination constraints from astrometry. As a result, we do not find any convincing evidence for mass-dependent kicks within the BH population, in agreement with \citet{zhao_evidence_2023}.

\citet{zhao_evidence_2023} find an anti-correlation of kick velocity with total binary mass, but also find no statistically significant difference between the natal kick distributions of BH and NS binaries. With our sample size, which is smaller than the one considered in \citet{zhao_evidence_2023}, it is unclear whether BH natal kicks are drawn from a similar distribution as NS natal kicks, or whether they are reduced (e.g., by a factor of $M_{\text{NS}} / M_{\text{BH}}$). We compare the BH systems considered in this work against NS systems with low-mass luminous companions in more detail in Appendix~\ref{appendix:co_comparison}. 

\subsection{Caveats}
\label{sec:caveats}

To compute the natal kick at birth (rather than at present-day), many studies in the literature integrate orbits back in time through the Galactic potential to $z = 0$ before computing the natal kick \citep[e.g.,][]{mandel_estimates_2016, atri_potential_2019, zhao_evidence_2023}. This accounts for the fact that these systems were most likely born closer to the disk midplane than they are today, as they have ages on the order of Gyr. Indeed, it is reasonable to assume that most LMXBs are several Gyrs old; we know that Gaia BH1 and Gaia BH2 are at least 5 Gyr old from the properties of their luminous stars \citep{andrae_el-badry_2023}, and analysis of the tertiary companion constrains V404 Cyg to be $3$--$5$ Gyr old \citep{burdge_2024}. However, this approach does not account for the fact that these systems have been dynamically heated by other processes than simply their natal kick over the course of their lifetimes. Due to these competing considerations, performing the natal kick calculation at the current spatial location is no worse than performing it at $z = 0$ \citep[for a study that finds that these approaches are consistent, see][]{zhao_evidence_2023}. Furthermore, except for the high-mass X-ray binary Cyg X-1, the ages of these binaries are poorly constrained. As such, it is impossible to integrate their trajectories back to their birth; instead, it is only possible to integrate back to a previous disk crossing, which is generally not the same point in time. It is important to note that BH kicks simply cannot be measured precisely in some cases. Broadly speaking, if the background stellar population has a large velocity dispersion, then we can neither prove nor rule out a natal kick.

In our analysis, we construct comparison samples at the present-day locations of each target. To test the effect of this assumption, we repeat our analysis, integrating the orbits of each system back to their last disk crossings and constructing comparison samples at those locations. Three systems (Swift J1727.8-162, MAXI J1305-704, and XTE J1118+480) are too far away from the present-day location of the Sun at their last disk crossing, rendering further kinematic analysis based on \textit{Gaia} DR3 infeasible. Of the remaining systems, six show a slight decrease in peculiar velocity and three show a slight increase in peculiar velocity. This translates into the minimum inferred kick velocity slightly increasing for three systems, slightly decreasing for two systems, and remaining unchanged for all other systems, with the changes being on the order of $\sim 10$ km s$^{-1}$. We note that this analysis is complicated by contamination from bulge stars when disk crossing locations are located far from the solar neighborhood.

An unknown fraction of stellar-mass BHs are in binaries. Since strong natal kicks are expected to unbind binaries, we expect BHs in binaries to preferentially have weaker kicks than isolated BHs, for which we currently have no useful constraints (as only one BH microlensing event has been identified to date). We attempt to quantify this bias by using the prescription of \citet{brandt_effects_1995} to probe whether BHs in binaries remain bound to their luminous companions following the supernova of the BH progenitor. We assume a pre-kick orbital period of $1$ day, a luminous companion mass of $1\,M_{\odot}$, a BH progenitor mass of $20\,M_{\odot}$, and a BH remnant mass of $10\,M_{\odot}$. We draw kick velocities from a Maxwellian distribution with $\sigma = 100$ km s$^{-1}$. At each kick velocity, we assign a random kick orientation and check whether the binary remains bound after the supernova. We repeat this simulation 100 times. We find that, while the true average kick is about $160$ km s$^{-1}$, the average kick of BHs that remain bound to their low-mass luminous companions is reduced to $142.4 \pm 0.7$ km s$^{-1}$. This translates to an $\approx 11$\% decrease in the average natal kick inferred from observations. We find that the average inferred natal kick is $150.0 \pm 0.3$ km s$^{-1}$ (percent decrease of $\approx 6\%$) and $157.6 \pm 0.2$ km s$^{-1}$ (percent decrease of $\approx 1\%$) if we assume lower progenitor masses of $15\,M_{\odot}$ and $10\,M_{\odot}$, respectively. On the other hand, if we increase the progenitor mass such that more than half of its mass is lost in the supernova, then a non-zero natal kick is actually required to keep the binary from becoming unbound.

Our inference ignores the fact that the velocity of the BHs is reduced by the presence of their companions due to conservation of momentum. Due to the generally large mass ratios of the systems we consider in this work, this is a small effect, except in the case of Cyg X-1. For a discussion of the impact of this effect on NS systems with low-mass companions, see Appendix \ref{appendix:co_comparison}.

\subsection{Implications for Theoretical Models}

In light of our results, simulations of core-collapse supernovae need to be able to explain why some BHs are born with a substantial natal kick, while others are born with virtually no kick at all. Indeed, the most compelling cases for both strong and weak natal kicks (Swift J1727.8-162's large space velocity and V404 Cyg's tertiary, respectively) were not known at the time of any previous analyses.

\citet{burrows_theory_2024} find two channels of black hole formation, one which leads to $\sim8$--$11\,M_{\odot}$ BHs that experience low natal kicks of $\lesssim 10$ km s$^{-1}$ (due to a failed supernova) and one which leads to $\sim2.5$--$3.5\,M_{\odot}$ BHs that experience significant recoil kicks of hundreds of km s$^{-1}$ (due to anisotropic accretion and neutrino-driven jets) strong enough to potentially unbind the binary. Similarly, \citet{janka_interplay_2024} find that BHs born via failed supernovae will experience small kicks on the order of $\lesssim$ few km s$^{-1}$, a conclusion reinforced by recent observational studies of VFTS 243 \citep{vigna-gomez_constraints_2024} and V404 Cyg \citep{burdge_2024}. \citet{janka_interplay_2024} additionally suggest that fallback supernovae in which the fallback mass is large enough to cause accretion-induced collapse of the proto-NS into a BH, but also small enough to allow the remainder of the star to successfully explode, could explain BHs that are born with a large natal kick.

The discussion above reinforces the theoretical expectation for a bimodal BH kick distribution. This evidence has important implications for binary evolution models, whose output parameters can depend sensitively on input assumptions about the magnitude of BH natal kicks. This, in turn, influences the predictions of binary population synthesis codes that rely on BH natal kick prescriptions to accurately model compact object populations and their progenitors \citep[see e.g.,][]{breivik_2020}. 

To illustrate, \citet{wysocki_2018} use isolated binary evolution models with different assumptions about natal kicks to explain LIGO's observations of gravitational wave (GW) events. They model BH natal kicks as being drawn from a 1D Maxwellian velocity distribution. Since kicks that are too large disrupt too many binaries and kicks that are too small cannot reproduce the observed range of spin-orbit misalignments, they estimate that BHs receive natal kicks on the order of 200 (50) km s$^{-1}$ if tidal processes do (not) realign stellar spins. \citet{boesky_2024} use \texttt{COMPAS} \citep[][]{compas_2022} to analyze the impact of the assumed 1D root-mean-square BH natal kick velocity on the GW merger rate, finding that a Maxwellian kick distribution with $\sigma = 265$ km s$^{-1}$ tends to reduce the observed rate by a factor of at least a few relative to a distribution with $\sigma = 30$ km s$^{-1}$. Further constraining the observed kick distribution is thus critical for modeling the progenitors of GW events. 

\section{Conclusion} \label{sec:conclusion}

We have constrained the natal kicks of a sample of dynamically confirmed stellar-mass black holes (BHs) in the Galactic disk with luminous companions (LCs) that have well-measured distances, proper motions, and systemic velocities (Table~\ref{tab:catalog}). We summarize our results below.

\begin{figure*}
\epsscale{0.8}
\plotone{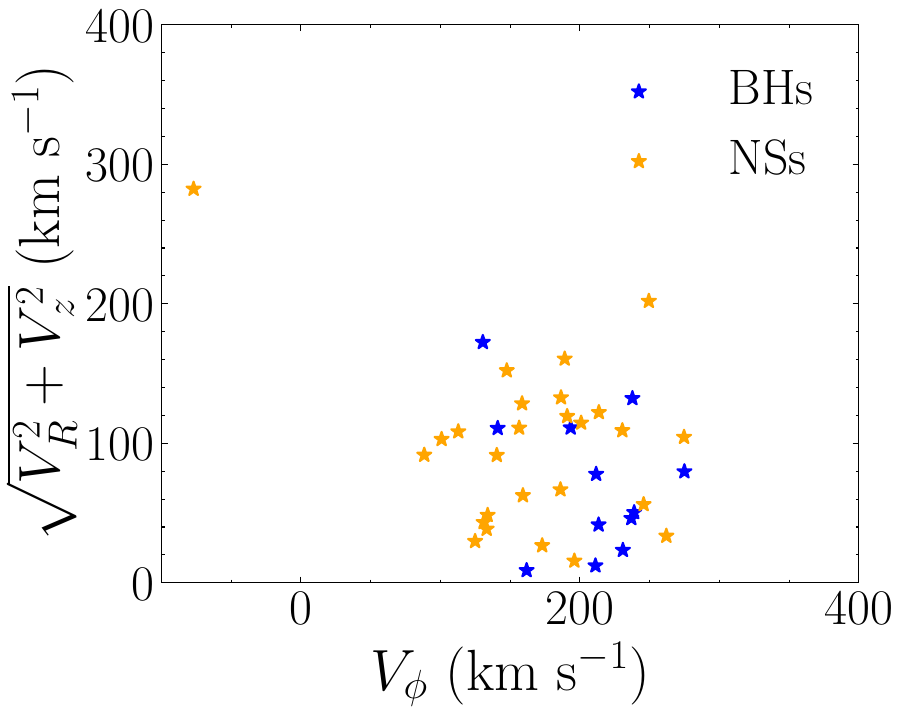}
\caption{Toomre diagram comparing kinematics of luminous companions to BHs (blue) and NSs (orange), respectively. We show both the BHs considered in this work and NSs from \citet{zhao_evidence_2023} with low-mass companions that have robust 6D measurements and are not in the Galactic halo or bulge. The NS systems are more dispersed than the BH systems, but not dramatically.}
\label{fig:co_comparison}
\end{figure*}

\begin{itemize}
    \item We construct Toomre diagrams for each of these systems (Figures~\ref{fig:kick_toomre_plots} and \ref{fig:no_kick_toomre_plots}). After accounting for the velocity dispersion of their local stellar populations, we find that half of these systems have higher peculiar velocities than 68\% of sources in their local comparison samples from \textit{Gaia} DR3, implying at least weak evidence for a natal kick (Table~\ref{tab:kicks}). The remainder of the systems have Toomre diagrams that are consistent with no natal kick.
    \item We integrate the orbits of these systems in the Galactic potential and show the results in Figure~\ref{fig:all_galactic_orbits}. The orbits of the systems hosting BHs that show evidence for a natal kick are generally puffier and more eccentric than the solar orbit, while the orbits of the systems hosting BHs that show no evidence of a natal kick are generally close to circular and do not travel far from the disk midplane, reinforcing our conclusions above.
    \item Comparison of our results to BH natal kicks derived in the literature shows that ignoring the local velocity dispersion tends to overestimate the magnitude of the inferred natal kick. Nevertheless, we find that two BH-LC systems (Swift J1727.8-162 and GRO J1655-40) show strong evidence for a BH natal kick, as their peculiar velocities are greater than those of 95\% of the local stellar population. 
    \item The unique orbits of V404 Cyg, which is part of a hierarchical triple, and VFTS 243, which is nearly circular despite not being tidally synchronized, allow for tighter kick constraints than kinematics alone. The observed properties of these systems strongly imply that their BHs received little to no natal kick. Combined with results from recent 3D simulations of core-collapse supernovae, we conclude that there is compelling observational and theoretical evidence that some BHs receive substantial natal kicks, while others receive almost no kick when they form. 
\end{itemize}

Selection effects and our small sample size limit inferences that can be made about the distribution of natal kicks of the entire Galactic population of stellar-mass BHs in binaries. Future discoveries, and further spectroscopic follow-up of systems hosting unconfirmed BH candidates, are necessary to tighten the constraints derived in this work.

\begin{acknowledgments}

We thank the referee for constructive feedback. We thank Tom Wagg and Lucas de S\'a for useful discussion. This research was supported by NSF grant AST-2307232. This work has made use of data from the European Space Agency (ESA) mission
{\it Gaia} (\url{https://www.cosmos.esa.int/gaia}), processed by the {\it Gaia}
Data Processing and Analysis Consortium (DPAC,
\url{https://www.cosmos.esa.int/web/gaia/dpac/consortium}). Funding for the DPAC
has been provided by national institutions, in particular the institutions
participating in the {\it Gaia} Multilateral Agreement.

\end{acknowledgments}

%



\software{astropy \citep{2013A&A...558A..33A,2018AJ....156..123A}, \texttt{gala} \citep{adrian_price_whelan_2024_10449846}}



\appendix

\section{Comparison of Black Hole Kicks to Neutron Star Kicks}
\label{appendix:co_comparison}

In Figure~\ref{fig:co_comparison}, we show a Toomre diagram summarizing the kinematics of luminous companions to BHs (blue) and NSs (orange). We compute velocity components in a Galactocentric frame as detailed in Section~\ref{sec:analysis}. We take parallaxes, proper motions, and systemic velocities for BH systems from Table~\ref{tab:catalog} and NS systems from the catalog compiled by \citet{zhao_evidence_2023}, respectively. We only include NS systems where robust 6D  measurements are available and the companion is a low-mass star, since NSs with high-mass companions will be dramatically slowed by the inertia of their companions. We do not include systems suspected to have been born in the halo, such as PSR J1024-0719 \citep{kaplan_2016} and Gaia NS1 \citep{el-badry_19_2024}. We also remove systems in the Galactic bulge. 

We find that the NS systems in Figure \ref{fig:co_comparison} are more dispersed than the BH systems, but not dramatically. Indeed, the dispersion of the NS systems in Figure \ref{fig:co_comparison} is less than the $\sigma = 265$ km s$^{-1}$ expected for isolated pulsars based on \citet{hobbs_statistical_2005}. This is primarily because conservation of momentum reduces the observed space velocities of NSs in binaries (where the companion is carried along with the NS) relative to isolated NSs, and also because the NSs in binaries born with the fastest kicks are unlikely to remain bound. Since BHs have higher masses than NSs, the former effect is less important for BHs with low-mass companions. 


\bibliography{bibliography}{}
\bibliographystyle{aasjournal}



\end{document}